\documentclass{article}

\usepackage{microtype}
\usepackage{amsfonts}
\usepackage{graphicx}
\usepackage {amsmath}
\usepackage{booktabs} 

\usepackage{hyperref}
\usepackage{colortbl}
\usepackage{subfig}
\usepackage{times,graphicx,amsmath,amssymb,booktabs,tabulary,multirow,overpic,xcolor}

\definecolor{sh_gray}{rgb}{0.84,0.84,0.84}
\definecolor{sh_gray2}{rgb}{1,0.89,0.75}
\definecolor{color3}{rgb}{0.95,0.95,0.95}
\definecolor{color4}{rgb}{0.96,0.96,0.86}
\definecolor{color5}{rgb}{0.90,0.90,0.90}

\newcolumntype{x}[1]{>{\centering\arraybackslash}p{#1pt}}

\newlength\savewidth
\newcommand{\tablestyle}[2]{\setlength{\tabcolsep}{#1}\renewcommand{\arraystretch}{#2}\centering\footnotesize}
\makeatletter\renewcommand\paragraph{\@startsection{paragraph}{4}{\z@}
	{.5em \@plus1ex \@minus.2ex}{-.5em}{\normalfont\normalsize\bfseries}}\makeatother

\usepackage[accepted]{icml2022}

\icmltitlerunning{Flow-Guided Sparse Transformer for Video Deblurring}

\begin{document}

	\twocolumn[
	\icmltitle{Flow-Guided Sparse Transformer for Video Deblurring}
	
	
	
	\icmlsetsymbol{equal}{*}
	
	\begin{icmlauthorlist}
		\icmlauthor{Jing Lin}{equal,thu}
		\icmlauthor{Yuanhao Cai}{equal,thu}
		\icmlauthor{Xiaowan Hu}{thu}
		\icmlauthor{Haoqian Wang$^{\dagger}$}{thu} 
		\icmlauthor{Youliang Yan}{huawei} \\
		\icmlauthor{Xueyi Zou$^{\dagger}$}{huawei}
		\icmlauthor{Henghui Ding}{ethz}
		\icmlauthor{Yulun Zhang}{ethz}
		\icmlauthor{Radu Timofte}{ethz}
		\icmlauthor{Luc Van Gool}{ethz}
	\end{icmlauthorlist}
	
	\icmlaffiliation{thu}{Shenzhen International Graduate School, Tsinghua University}
	\icmlaffiliation{huawei}{Huawei Noah's Ark Lab}
	\icmlaffiliation{ethz}{ETH Z\"{u}rich}
	
	\icmlcorrespondingauthor{Haoqian Wang}{wanghaoqian@tsinghua.edu.cn}
	\icmlcorrespondingauthor{Xueyi Zou}{zouxueyi@huawei.com}
	
	\icmlkeywords{Machine Learning, ICML}
	
	\vskip 0.3in
	]
	
	
	
	\printAffiliationsAndNotice{\icmlEqualContribution} 
	
	\begin{figure*}[t] 
		\begin{center} \hspace{-0mm}
			\begin{tabular}[t]{c}
				\includegraphics[width=0.95\textwidth]{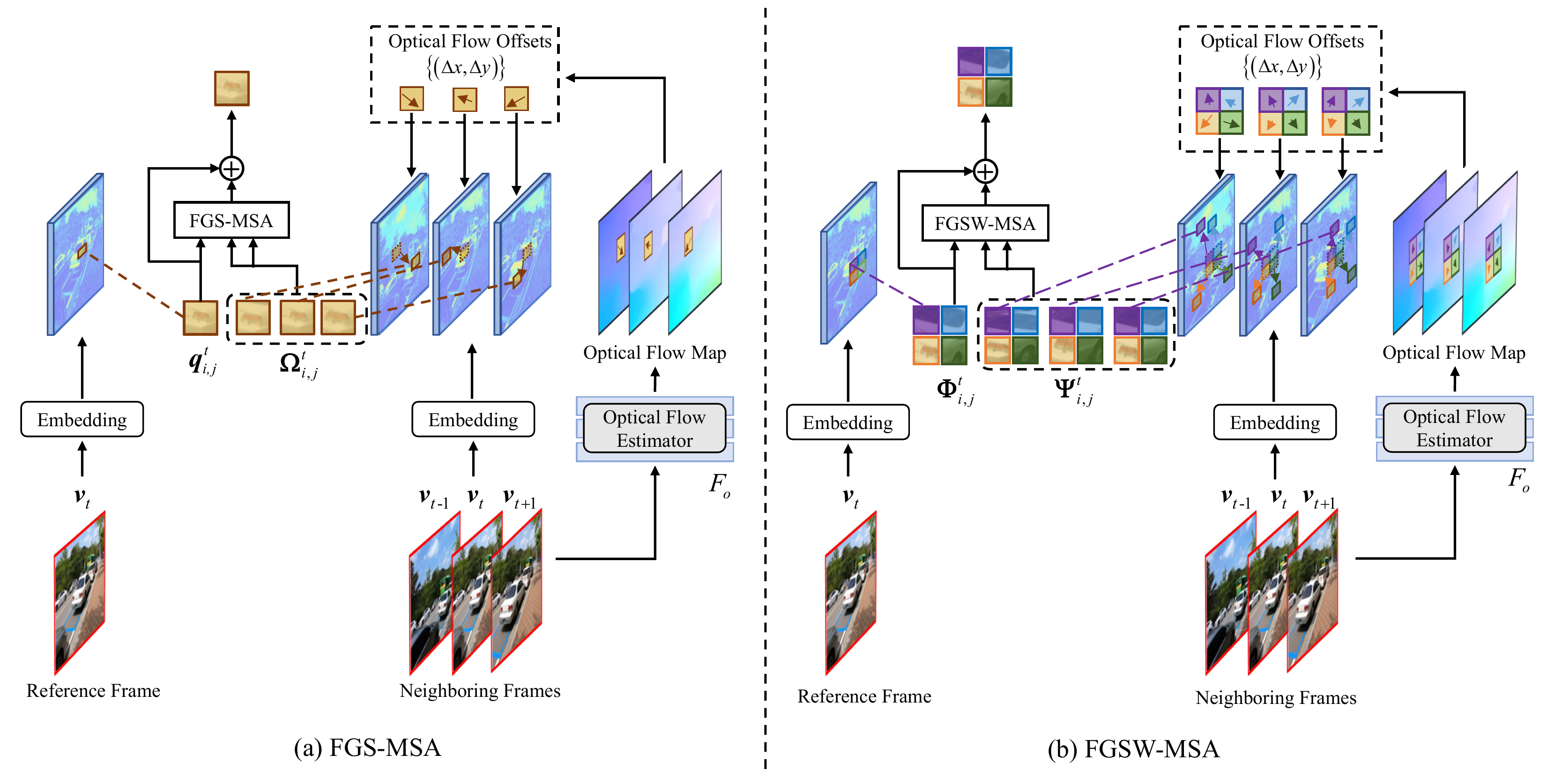}
			\end{tabular}
		\end{center}
		\vspace*{-6mm}
		\caption{\small The illustration of our flow-guided self-attention mechanisms. (a) FGS-MSA globally samples spatially sparse yet highly related $key$ elements of similar and sharper patches in neighboring frames.  (b)  Instead of sampling a single $key$ element on each neighboring frame,  FGSW-MSA robustly samples all the $key$ elements corresponding to the $query$ elements of the window on the reference frame.}
		\label{fig:teaser}
		\vspace{-1mm}
	\end{figure*}

	\begin{abstract}
		\vspace{-1mm}
		Exploiting similar and sharper scene patches in spatio-temporal neighborhoods is critical for video deblurring. However, CNN-based methods show limitations in capturing long-range dependencies and modeling non-local self-similarity. In this paper, we propose a novel framework, Flow-Guided Sparse Transformer (FGST), for video deblurring. In FGST, we customize a self-attention module, Flow-Guided Sparse Window-based Multi-head Self-Attention (FGSW-MSA). For each $query$ element on the blurry reference frame, FGSW-MSA enjoys the guidance of the estimated optical flow to globally sample spatially sparse yet highly related $key$ elements corresponding to the same scene patch in neighboring frames. Besides, we present a Recurrent Embedding (RE) mechanism to transfer information from past frames and strengthen long-range temporal dependencies. Comprehensive experiments demonstrate that our proposed FGST outperforms state-of-the-art (SOTA) methods on both DVD and GOPRO datasets and yields visually pleasant results in real video deblurring.  \url{https://github.com/linjing7/VR-Baseline}
	\end{abstract}
	
	\vspace{-4mm}
	\section{Introduction}
	\vspace{-1mm}
	\label{introduction}
	
	Video deblurring is a fundamental yet challenging task in low-level computer vision and graphics communities. It aims to restore the latent frames from a blurry video sequence. Serving as a preprocessing technique, video deblurring has wide applications such as video stabilization~\cite{matsushita2006full}, tracking~\cite{track}, autonomous driving~\cite{3d_det}, \emph{etc.}  Hand-held devices are more and more popular in capturing videos of dynamic scenes, where prevalent depth variations, abrupt camera shakes, and high-speed object movements lead to undesirable blur in videos. To alleviate the effect of motion blur, researchers have put a lot of efforts into video deblurring.  
	
	Conventional methods are mainly based on hand-crafted priors and assumptions, which limits the model capacity. Besides, the assumptions on motion blur and latent frames usually lead to complex energy functions that are difficult to solve. Also, the inaccurately estimated motion blur kernel with hand-crafted priors may easily result in severe artifacts.
	
	In the past decade, video deblurring has witnessed significant progresses with the development of deep learning. Convolutional neural network (CNN) applies a powerful model to learn the mapping from blurry videos to sharp videos under the supervision of a large-scale dataset of blurry-sharp video pairs. CNN-based methods yield impressive performance but show limitations in modeling long-range spatial dependencies and capturing non-local self-similarity.
	
	Recently, the emergence of Transformer provides an alternative to alleviate the constraints of CNN-based methods. \textbf{Firstly}, Transformer excels at modeling long-range spatial dependencies. The contextual information and spatial correlations are critical to restoring the motion blur. \textbf{Secondly}, similar and sharper scene patches from neighboring frames provide crucial cues for video deblurring. Fortunately, the self-attention module in Transformer is dedicated to calculating the correlations among pixels and capturing the self-similarity along the temporal sequence. Thus, Transformer inherently resonates with the goal of learning similar information from spatio-temporal neighborhoods. \textbf{Nevertheless}, directly using existing  Transformers for video deblurring has two issues. \textbf{On one hand}, when the standard global Transformer~\cite{global_msa} is utilized, the computational cost is quadratic to the spatio-temporal dimensions. This burden is nontrivial and sometimes unaffordable. Meanwhile, the global Transformer attends to redundant $key$ elements, which may easily cause non-convergence issue~\cite{de_detr} and over-smoothing results~\cite{xiangtl_gald}. \textbf{On the other hand}, when the local window-based Transformer~\cite{liu2021swin} is used, the self-attention is calculated within position-specific windows, causing limited receptive fields. The model may neglect some $key$ elements of similar and sharper scene patches in the spatio-temporal neighborhood when fast motions are present. We summarize the main reason for the above problems, \emph{i.e.}, previous Transformers lack the guidance of motion information, when calculating self-attention. We note that the motion information can be estimated by optical flow.
	
	Exploiting an optical flow estimator to capture motion information and align neighboring frames is a common strategy in video restoration~\cite{makansi2017end,Su,xue2019video,tsp}. Previous flow-based methods mainly adopt the pre-warping strategy. Specifically, they employ an optical flow estimator to produce motion offsets, warp neighboring frames, and align regions corresponding to the same object but misaligned in neighboring image or feature domains. This scheme suffers from the following issues: \textbf{(i)} The interpolating operations in the warping module modify the original image information. As a result, some critical image priors such as self-similarity and sharp textures may be sacrificed. Undesirable artifacts may be introduced to the restored video and the deblurring performance may degrade. \textbf{(ii)} The frame alignment and subsequent representation aggregation are separated. This paradigm is inflexible and does not make full use of optical flow. Besides, the deblurring results are easily affected by the performance of the optical flow estimator. The robustness of this scheme can be further improved.
	
	\begin{figure*}[t]
		\begin{center}
			\begin{tabular}[t]{c} \hspace{-2mm}
				\includegraphics[width=0.95\textwidth]{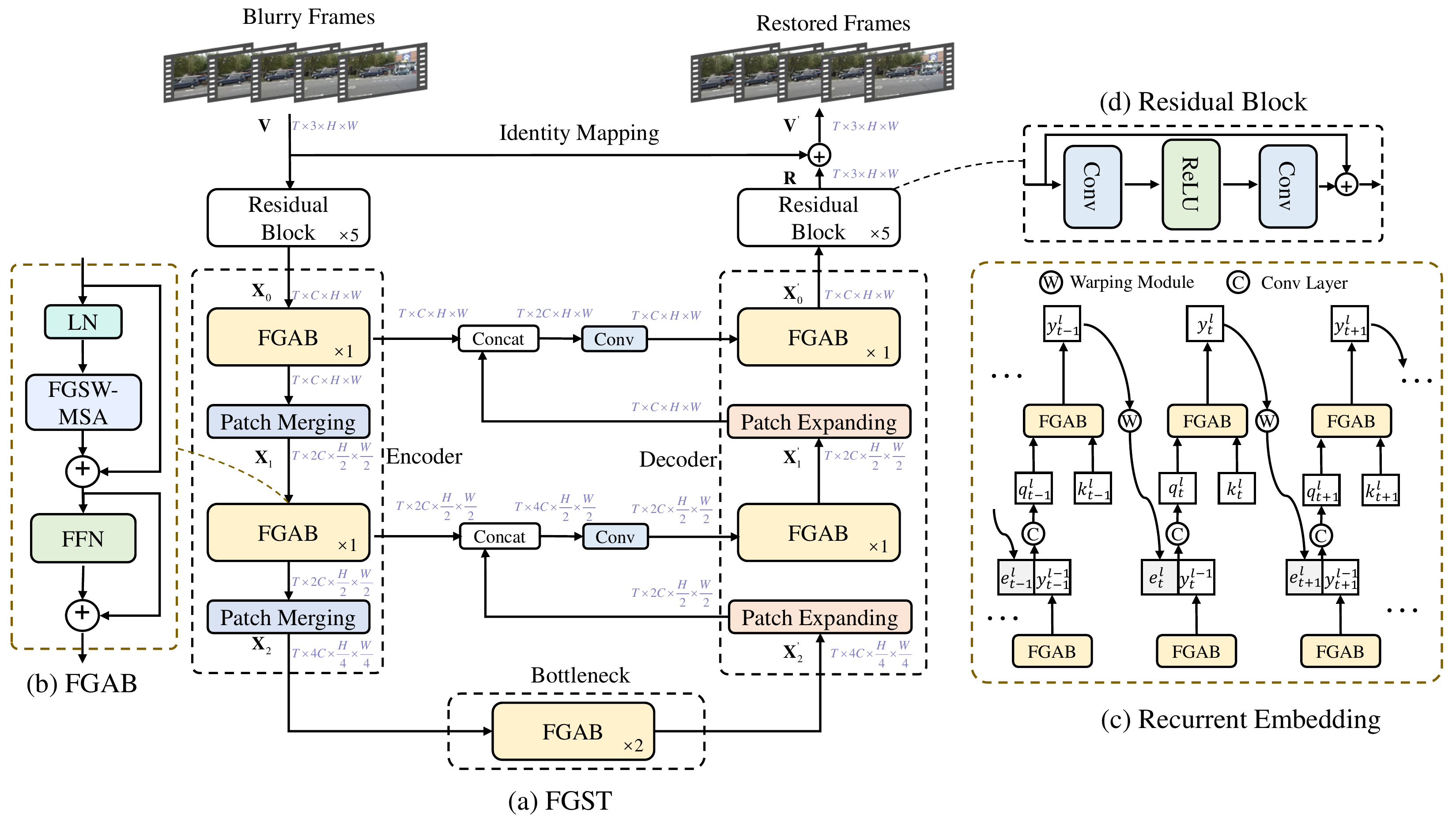}
			\end{tabular}
		\end{center}
		\vspace*{-6.5mm}
		\caption{\small The architecture of FGST. (a) FGST consists of an encoder, a bottleneck, and a decoder. FGST is built up by FGABs. (b) FGAB is composed of a layer normalization, an FGSW-MSA, and a feed-forward network. (c) RE aggregates the output of the last frame and the input of the current frame. Some intermediate steps between FGABs are omitted. (d) The components of residual block.}
		\label{fig:pipeline}
		\vspace{-1mm}
	\end{figure*}
	
	This work aims to cope with the above problems. We propose a novel method, Flow-Guided Sparse Transformer (FGST), for video deblurring. \textbf{Firstly}, we adopt Transformer instead of CNN as the deblurring model because of its advantages of capturing long-range spatial dependencies and non-local self-similarity. \textbf{Secondly}, to alleviate the limitations of previous Transformers and the pre-warping strategy, we customize Flow-Guided Sparse Multi-head Self-Attention (FGS-MSA) as shown in Fig.~\ref{fig:teaser} (a). For each $query$ element on the reference frame, FGS-MSA guided by an optical flow estimator globally samples spatially sparse $key$ elements corresponding to the same scene patch but misaligned in the neighboring frames. These sampled $key$ elements provide self-similar and highly related image prior information, which is critical to restoring motion blur. Different from original global and local Transformers, our FGST neither blindly samples redundant $key$ elements nor suffers from limited receptive fields. Meanwhile, our alignment scheme is different from the pre-warping operation mainly used by previous flow-based methods. Instead of warping the neighboring frames, our FGST samples $key$ elements in consecutive frames to calculate the self-attention. Thus, the original image prior information can be preserved. \textbf{Thirdly}, we  promote FGS-MSA to Flow-Guided Sparse Window-based Multi-head Self-Attention (FGSW-MSA) as shown in Fig.~\ref{fig:teaser} (b). The feature maps are split into non-overlapping windows. Instead of sampling a single $key$ element on each neighboring frame for a single $query$ element, FGSW-MSA samples $key$ elements assigned by the optical flow corresponding to all the $query$ elements of the window on the reference frame. Thus, FGSW-MSA is more robust to accommodate pixel-level flow offset prediction deviations. \textbf{Finally}, our FGSW-MSA is calculated within a short temporal sequence reducing the computational cost. Hence, the receptive field of FGSW-MSA is spatially global but temporally local. Motivated by RNN-based methods~\cite{Nah,RNN_3}, we propose Recurrent Embedding (RE) to transfer information of past frames and capture long-range temporal dependencies.
	
	Our contributions can be summarized as follows:
	
	\begin{itemize}
		\setlength{\itemsep}{1pt}
		\setlength{\parsep}{1pt}
		\setlength{\parskip}{1pt}
		\vspace{-4mm}
		\item We propose a new method, FGST, for video deblurring. To the best of our knowledge, it's the first attempt to explore the potential of Transformer in this task. 
		\item We customize a novel self-attention mechanism, FGS-MSA, and its improved version, FGSW-MSA.
		\item We design an embedding scheme, RE, to transfer frame information and capture temporal dependencies.
		\item Our FGST outperforms SOTA methods on DVD and GOPRO datasets by a large margin and yields more visually pleasing results in real-world video deblurring. 
	\end{itemize}
	
	\section{Related Work}
	\label{related_work}
	\subsection{Video Deblurring}
	In recent years, the deblurring research focus is shifting from single image deblurring~\cite{zoran2011learning,chakrabarti2016neural,purohit2020region} to the more challenging video deblurring~\cite{real_blur,matsushita2006full}. Traditional methods~\cite{li2010generating,zhang2013multi} are based on hand-crafted image priors and assumptions, which lead to limited generality and representing capacity. With the development of deep learning, recent methods are mainly CNN-based or RNN-based. \cite{dblrnet} employ 3D convolutions to model spatio-temporal relations of frames. \cite{hyun2017online} and \cite{Nah} use RNN-based models to restore the latent frames. However, CNN-based methods show limitations in capturing long-range dependencies while RNN-based methods are not sensitive to patch-level spatial correlation and  motion information. 
	
	\subsection{Vision Transformer}
	Transformer is firstly proposed by \cite{vaswani2017attention} for machine translation. Recently, Transformer has been introduced to high-level~\cite{global_msa,liu2021swin,de_detr,SETR,xcit,to_1,prtr,tc_2,tc_3,rsn} and low-level vision~\cite{ipt,cai2021mask,uformer,vsrt,cai2021learning,hu2021pseudo}. \cite{arnab2021vivit} factorize the spatial and temporal dimensions of the input video and propose a Transformer model for video classification. \cite{ipt} present a large model IPT pre-trained on large-scale datasets with a multi-task learning scheme. \cite{vsrt} propose VSR-Transformer that uses the self-attention mechanism for better feature fusion in video super-resolution, but image features are still extracted from CNN. \cite{uformer} use Swin Transformer~\cite{liu2021swin} blocks to build up a U-shaped structure for single image restoration.  In \cite{vaswani2021scaling,cao2021swin,liu2021swin}, window-based local self-attention is adopted to replace the global self-attention module of the standard Transformer. However, directly using previous global or local Transformers for video deblurring leads to unaffordable computational cost or limited receptive fields. 
	
	\vspace{-1mm}
	\subsection{Flow-based Video Restoration}
	\vspace{-1mm}
	Optical flow estimators are widely used in video restoration tasks~\cite{gast2019deep,xue2019video,gong2017motion,sun2015learning,makansi2017end,Su,tsp} to align highly related but mis-aligned frames. Previous flow-based video deblurring  methods~\cite{xue2019video,makansi2017end,Su,tsp,gast2019deep} mainly adopt the pre-warping strategy, which firstly estimates the optical flow and then warps the neighboring frames. For example, \cite{Su} experiments with pre-warping input images based on classic optical flow methods to register them to the reference frame. Nonetheless, this flow-based pre-warping scheme separates the frame alignment and subsequent information aggregation. The original frame information is sacrificed and the guidance effect of optical flow is not fully explored. 
	
	\vspace{-1.5mm}
	\section{Method}
	\vspace{-0.5mm}
	\subsection{Overall Architecture}
	Figure~\ref{fig:pipeline} (a) shows the architecture of FGST that adopts the widely used U-shaped structure, consisting of an encoder, a bottleneck, and a decoder. Figure~\ref{fig:pipeline} (b) depicts the basic unit of FGST, $i.e.$, Flow-Guided Attention Block (FGAB). 
	
	The input is a blurry video $\mathbf{V}\in \mathbb{R}^{T\times 3\times H \times W}$, where $T$ denotes the sequence length, $H$ and $W$ denote the width and height of the frame. \textbf{Firstly}, FGST exploits 5 residual blocks to map $\mathbf{V}$ into tokens $\mathbf{X}_0\in \mathbb{R}^{T\times C\times H \times W}$, where $C$ denotes the channel number. The details of residual block are shown in Fig.~\ref{fig:pipeline} (d). \textbf{Secondly}, $\mathbf{X}_0$ passes through two FGABs and patch merging layers to generate hierarchical features. The patch merging layer is a strided 4$\times$4 convolution that downsamples the feature maps and doubles the channels. Thus, the  tokens of the $i_{th}$ layer in the encoder are denoted as $\mathbf{X}_i\in \mathbb{R}^{T\times 2^{i}C\times \frac{H}{2^{i}} \times \frac{W}{2^{i}}}$. \textbf{Thirdly}, $\mathbf{X}_2$ passes through the bottleneck, which consists of two FGABs.
	
	\textbf{Subsequently}, following the spirit of U-Net~\cite{unet}, we customize a symmetrical decoder, which is composed of two FGABs and patch expanding layers. The patch expanding layer is a strided 2$\times$2 deconvolution that upsamples the feature maps. To alleviate the information loss caused by downsampling, skip connections are used for feature fusion between the encoder and decoder.
	
	After undergoing the decoder, the feature maps pass through 5 residual blocks to generate a residual frame sequence $\mathbf{R}\in \mathbb{R}^{T\times 3\times H \times W}$. \textbf{Finally}, the deblurred video $\mathbf{V'}\in \mathbb{R}^{T\times 3\times H \times W}$ can be derived by $\mathbf{V'} = \mathbf{V} + \mathbf{R}$.
	
	\vspace{-1.5mm}
	\subsection{Flow-Guided Attention Block}
	\vspace{-1mm}
	As analyzed in Sec.~\ref{introduction}, the standard global Transformer brings quadratic computational complexity with respect to the token number and easily leads to non-convergence issue and over-smoothing results. The previous window-based local Transformers suffer from the limited receptive fields. 
	
	To address these problems, we propose to use optical flow as the guidance to sample $key$ elements from spatio-temporal neighborhoods when calculating the self-attention. Based on this motivation, we customize the basic unit, FGAB as shown in Fig.~\ref{fig:pipeline} (b). FGAB consists of a layer normalization (LN), a Flow-Guided Sparse Window-based Multi-head Self-Attention (FGSW-MSA), a feed-forward network (FFN), and two identity mappings. The FFN is composed of 5 consecutive residual blocks. In this part, we first introduce Flow-Guided Sparse Multi-head Self-Attention (FGS-MSA) and then its improved version, FGSW-MSA.
	
	\noindent\textbf{FGS-MSA.} The details of FGS-MSA are shown in Fig.~\ref{fig:teaser} (a). Given the $\boldsymbol{t}_{th}$ input blurry video frame $\boldsymbol{v}_t \in \mathbb{R}^{3\times H \times W}$ as the reference frame, $\boldsymbol{q}_{i,j}^t$ and $\boldsymbol{k}_{i,j}^t \in \mathbb{R}^{C}$  respectively denote the \emph{query} and \emph{key} elements at the position ($i$,$j$) on $\boldsymbol{v}_t$. FGS-MSA aims to model long-range spatial dependencies and capture non-local self-similarity. To this end, FGS-MSA produces $key$s from the $key$ elements of similar and sharper scene patches in the spatio-temporal neighborhood of $\boldsymbol{v}_t$. The $key$ sampling is directed by the motion information that is predicted by an optical flow estimator. This set of $key$ elements is corresponding to $\boldsymbol{q}_{i,j}^t$ and we denote it as 
	\begin{equation}
	\small
	\mathbf{\Omega}_{i,j}^t = \{\boldsymbol{k}^f_{i+{\Delta x_f},j+{\Delta y_f}}\bigm||f-t| \leq r\},
	\label{eq:Omega_k}
	\end{equation}
	where $r$ represents the temporal radius of the neighboring frames. $({\Delta x_f}, {\Delta y_f})$ denotes the value at position ($i, j$) of the estimated motion offset map, which is predicted from the reference frame $\boldsymbol{v}_t$ to the neighboring frame $\boldsymbol{v}_f$:
	\begin{equation}
	\small
	({\Delta x_f}, {\Delta y_f}) = [F_o(\boldsymbol{v}_t, \boldsymbol{v}_f)~(i,j)],
	\label{eq:flow}
	\end{equation}
	where $F_o$ denotes the mapping function of the optical flow estimator and [$\cdot$] refers to the rounding operation. Subsequently, FGS-MSA can be formulated as
	\vspace{-0.5mm}
	\begin{equation}
	\small
	\text{FGS-MSA}(\boldsymbol{q}_{i,j}^t,\mathbf{\Omega}_{i,j}^t) = \sum_{n=1}^{N} \mathbf{W}_n  \sum_{\boldsymbol{k}\in\mathbf{\Omega}_{i,j}^t} \mathbf{A}_{n\boldsymbol{q}_{i,j}^t\boldsymbol{k}}~ \mathbf{W'}_n ~  \boldsymbol{k},
	\label{eq:OFGMSA}
	\end{equation}
	where $N$ is the number of the attention heads. $\mathbf{W}_n \in \mathbb{R}^{C\times d}$ and $\mathbf{W'}_n \in \mathbb{R}^{d\times C}$ are learnable parameters, where $d = \frac{C}{N}$ denotes the representation dimension per head.  $\mathbf{A}_{n\boldsymbol{q}_{i,j}^t\boldsymbol{k}}$ is the self-attention of the $n_{th}$ head, which is formulated as
	\vspace{-0.5mm}
	\begin{equation}
	\mathbf{A}_{n\boldsymbol{q}_{i,j}^t\boldsymbol{k}} = \underset{\boldsymbol{k}\in\mathbf{\Omega}_{i,j}^t}{\text{softmax}} (\frac{(\boldsymbol{q}_{i,j}^t)^T\mathbf{U}_n^T\mathbf{V}_n\boldsymbol{k}}{\sqrt{d}}),
	\label{eq:ScaledDotProductAttn}
	\end{equation}
	where $\mathbf{U}_n$ and $\mathbf{V}_n \in \mathbb{R}^{d\times C}$ are learnable parameters. Given an input $\mathbf{V} \in \mathbb{R}^{T\times 3 \times H \times W}$, the computational cost of the global MSA~\cite{global_msa} and FGS-MSA are
	\vspace{-1mm}
	\begin{equation}
	\small
	\begin{aligned}
	O{(\text{global MSA})}&=4(THW)C^2 + 2(THW)^2C,\\
	O{(\text{FGS-MSA})}&= 2(THW)C\big(2(r+1)C+2r+1\big).\\
	\label{eq:complexity}
	\end{aligned}
	\vspace{-6mm}
	\end{equation}
	The standard global MSA leads to quadratic ($(THW)^2$) computational complexity while our proposed FGS-MSA contributes to much cheaper linear computational cost with respect to the token number $(THW)$. Detailed analysis are provided in the supplementary material (SM).
	
	
	\noindent\textbf{FGSW-MSA.} For each neighboring frame, FGS-MSA only samples a single $key$ element. When the optical flow estimation is inaccurate, the deblurring performance may be easily affected. To further improve the robustness and reliability of our method, we promote FGS-MSA to FGSW-MSA. As shown in Fig.~\ref{fig:teaser} (b), the feature maps are split into non-overlapping windows. The spatial size of each window is $M\times M$. $\mathbf{\Phi}_{i,j}^t$ denotes the set of $query$ elements in the window centering at position $(i,j)$ of the $t_{th}$ frame:
	\vspace{-1mm}
	\begin{equation}
	\small
	\mathbf{\Phi}_{i,j}^t = \{\boldsymbol{q}_{m,n}^t \big|~|m-i| \leq M/2, |n-j| \leq M/2\}.
	\label{eq:FGSWMSA_1}
	\vspace{-1mm}
	\end{equation} 
	For each $\boldsymbol{q}_{m,n}^t \in \mathbf{\Phi}_{i,j}^t$, FGSW-MSA samples not only its corresponding $key$ elements in $\mathbf{\Omega}_{m,n}^t$ (Eq.~\eqref{eq:Omega_k}) assigned by the flow offsets but also the $key$ elements corresponding to other $query$ elements in $\mathbf{\Phi}_{i,j}^t$. We denote the set of these $key$ elements as $\mathbf{\Psi}_{i,j}^t$, which can be formulated as
	\vspace{-1mm}
	\begin{equation}
	\small
	\mathbf{\Psi}_{i,j}^t = \underset{\footnotesize |m-i| \leq M/2,~|n-j| \leq M/2}{\bigcup~~\mathbf{\Omega}_{m,n}^t}.
	\label{eq:FGSWMSA_2}
	\vspace{-1mm}
	\end{equation}
	Instead of attending to a single $key$ element on each neighboring frame for a single $query$, FGSW-MSA pays attention to the $key$ elements from similar and sharper scene patches corresponding to all $query$ elements in $\mathbf{\Phi}_{i,j}^t$. The attending region is enlarged from pixel to window. Thus, FGSW-MSA is more robust to accommodate pixel-level flow prediction deviations. FGSW-MSA can be formulated as
	\vspace{-0.5mm}
	\begin{equation}
	\small
	\text{FGSW-MSA}(\mathbf{\Phi}_{i,j}^t,\mathbf{\Psi}_{i,j}^t) = \{ \text{FGS-MSA}(\boldsymbol{q},\mathbf{\Psi}_{i,j}^t) | \boldsymbol{q} \in \mathbf{\Phi}_{i,j}^t \}.
	\vspace{-0.5mm}
	\end{equation}
	Given the input $\mathbf{V}$, the computational complexity is
	\vspace{-0.5mm}
	\begin{equation}
	\small
	O(\text{FGSW-MSA}) = 2(THW)C\big(C+(2r+1)(C+M^2)\big).
	\label{eq:complexity_2}
	\vspace{-0.5mm}
	\end{equation}
	The computational cost of FGSW-MSA is linear with respect to the number of tokens ($THW$). Eq.~\eqref{eq:complexity} and \eqref{eq:complexity_2} reveal the high efficiency and resource economy of our FGST. Please refer to the SM for more detailed analysis.
	
	\begin{figure}[h]
		\begin{center}
			\begin{tabular}[t]{c} \hspace{-3mm} 
				\includegraphics[width=0.48\textwidth]{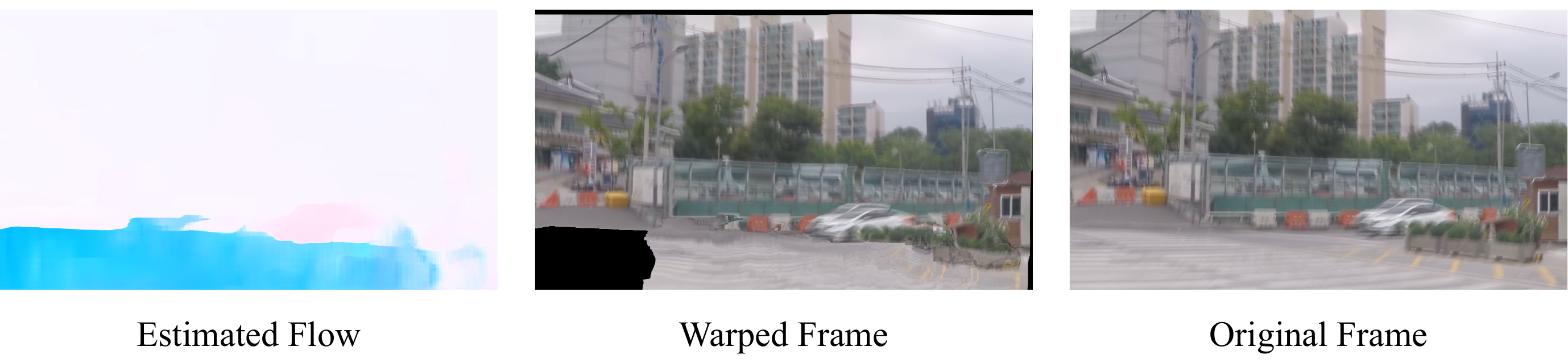}
			\end{tabular}
		\end{center}
		\vspace{-5mm}
		\caption{\small The pre-warping strategy mainly adopted by previous video deblurring methods  sacrificies the input image information.}
		\label{fig:pre_warping}
		\vspace{-3mm}
	\end{figure}
	
	\begin{figure*}[t]
		\begin{center}
			\begin{tabular}[t]{c} \hspace{-2mm}
				\includegraphics[width=1.0\textwidth]{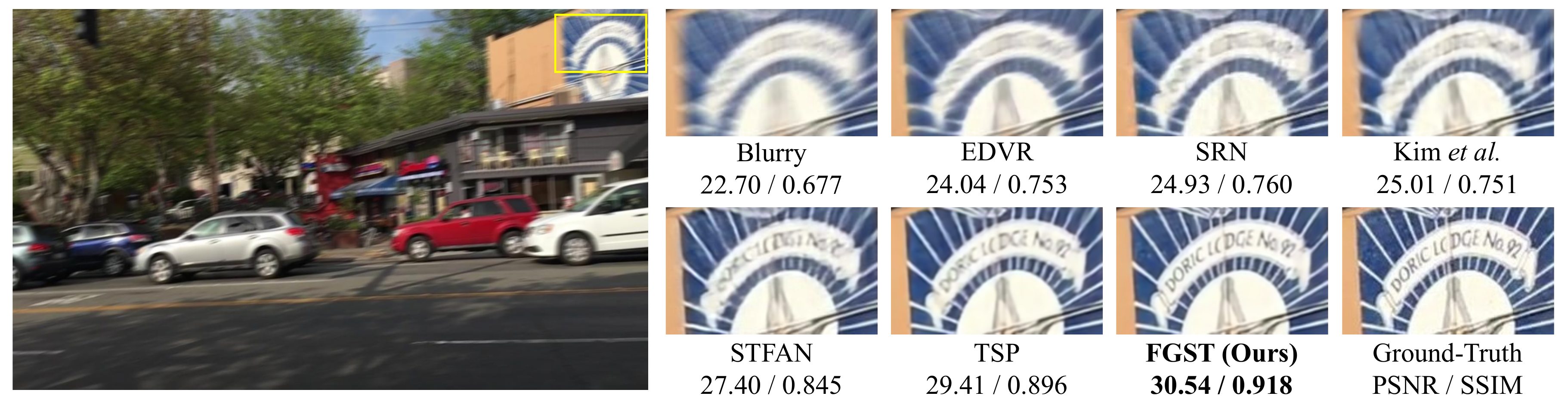}
			\end{tabular}
		\end{center}
		\vspace*{-5mm}
		\caption{\small Visual comparisons between FGST and SOTA methods on DVD dataset~\cite{Su}. Please zoom in for a better view.}
		\label{fig:dvd}
		\vspace{-3mm}
	\end{figure*}
	
	\begin{table*}[t]
		\begin{center}
			\setlength{\tabcolsep}{2.5pt}
			\scalebox{0.62}{
				\begin{tabular}{l c c c c c c c c c c}
					\toprule
					\rowcolor{color3} Method & Kim and Lee & Gong \emph{et al.}  & Su \emph{et al.} & Kim \emph{et al.}    & STFAN    & Xiang \emph{et al.}    & TSP    & Suin \emph{et al.}    & ARVo    & \textbf{FGST}   \\
					\rowcolor{color3} &\cite{Kim} &\cite{gong2017motion}  &\cite{Su} &\cite{hyun2017online} &\cite{stfan} &\cite{Xiang} &\cite{tsp} &\cite{Suin} &\cite{arvo} &\textbf{(Ours)} \\
					\midrule
					PSNR~$\textcolor{black}{\uparrow}$ &26.94 &28.27 &30.01 &29.95 &31.15 &31.68 &32.13 &32.53 &32.80 &\textbf{33.36}\\
					SSIM~$\textcolor{black}{\uparrow}$ &0.816 &0.846 &0.888 &0.869 &0.905 &0.916 &0.927 &0.947 &0.935 &\textbf{0.950}\\
					\bottomrule
			\end{tabular}}
			\vspace{-3mm}
			\caption{\small Video deblurring results  compared with other methods on the DVD benchmark \cite{Su}. FGST achieves SOTA results.}
			\label{tab:dvd}
		\end{center}\vspace{-4mm}
	\end{table*}
	
	\begin{figure*}[h]
		\begin{center}
			\begin{tabular}[t]{c} \hspace{-2.5mm}
				\includegraphics[width=\textwidth]{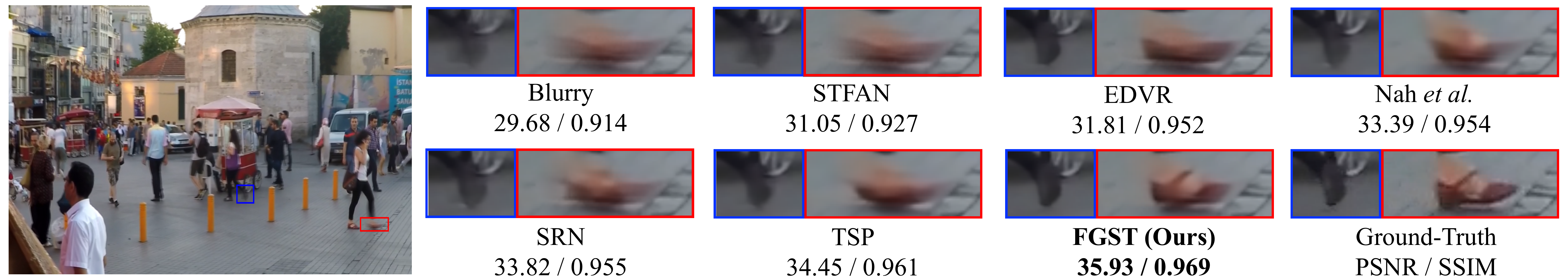}
			\end{tabular}
		\end{center}
		\vspace*{-4mm}
		\caption{\small Visual comparisons between our FGST and SOTA methods on GOPRO dataset~\cite{GoPro}. Zoom in for a better view.}
		\label{fig:gopro}
		\vspace{-3mm}
	\end{figure*}
	
	\begin{table*}[t]
		\begin{center}
			\setlength{\tabcolsep}{2.5pt}
			\scalebox{0.625}{
				\begin{tabular}{l c c c c c c c c c c c}
					\toprule
					\rowcolor{color3} Method  & Gong \emph{et al.} & Kim \emph{et al.}   &EDVR &Su \emph{et al.}   & STFAN    &Nah \emph{et al.}    &Tao \emph{et al.}    &TSP    & Suin \emph{et al.}    & \textbf{FGST}   \\
					\rowcolor{color3} &\cite{gong2017motion} &\cite{hyun2017online} &\cite{edvr} &\cite{Su} &\cite{stfan} &\cite{Nah} &\cite{Tao} &\cite{tsp} &\cite{Suin} &\textbf{(Ours)} \\
					\midrule
					PSNR~$\textcolor{black}{\uparrow}$ &26.06 &26.82 &26.83 &27.31 &28.59 &29.97 &30.29 &31.67 &32.10 &\textbf{32.90}\\
					SSIM~$\textcolor{black}{\uparrow}$ &0.863 &0.825 &0.843 &0.826 &0.861 &0.895 &0.901 &0.928 &0.960 &\textbf{0.961}\\
					\bottomrule
			\end{tabular}}
			\vspace{-4mm}
			\caption{\small Video deblurring results  compared with other methods on the GOPRO dataset \cite{GoPro}. FGST achieves SOTA results.}
			\label{tab:gopro}
		\end{center}\vspace{-5.5mm}
	\end{table*}
	
	\begin{figure}[h]
		\begin{center}
			\begin{tabular}[t]{c} \hspace{-2mm} 
				\includegraphics[width=0.48\textwidth]{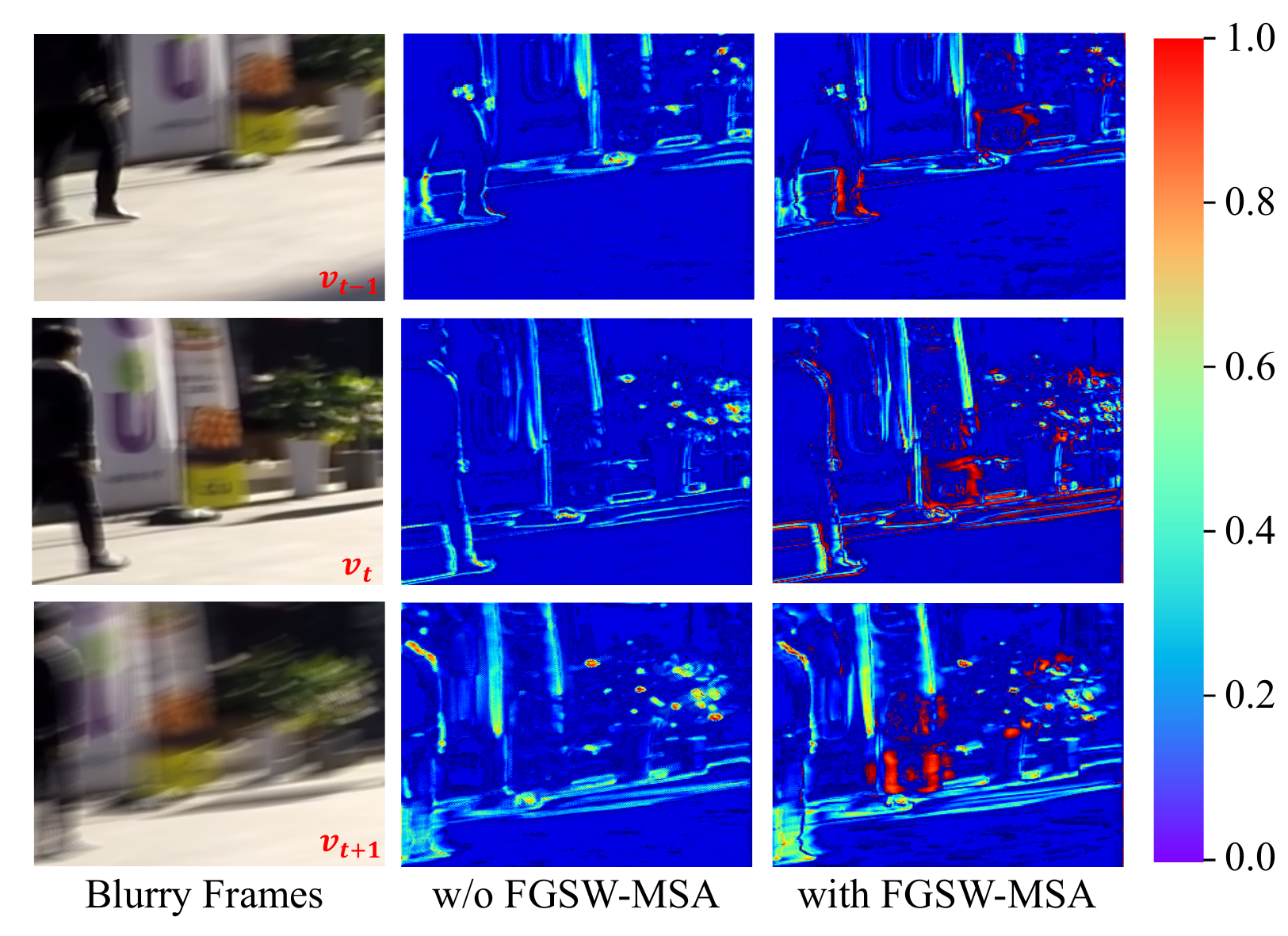}
			\end{tabular}
		\end{center}
		\vspace{-6mm}
		\caption{\small We visualize the last feature maps of the deblurring models with and without FGSW-MSA. The model using our FGSW-MSA pays more attention to similar but misaligned scene  patches.}
		\label{fig:fea}
		\vspace{-5mm}
	\end{figure}
	
	\noindent\textbf{Discussion.} \textbf{(i)} Our FGSW-MSA enjoys much larger receptive fields than W-MSA~\cite{liu2021swin}. Specifically, according to Eq.~\eqref{eq:Omega_k}, \eqref{eq:flow}, \eqref{eq:FGSWMSA_1}, and \eqref{eq:FGSWMSA_2}, the receptive field of FGSW-MSA can cover the whole input feature map when the estimated flow offset is large enough. In practice, the motion offset predicted by the optical flow estimator between two adjacent frames can reach 40 and 38 pixels on GOPRO and DVD datasets. The input spatial size is 256$\times$256. $M$ is set to 3. Thus, the receptive field of FGSW-MSA can reach 83$\times$83 (83 = 40$\times$2+3) and 79$\times$79 while that of W-MSA is still 3$\times$3. \textbf{(ii)} Unlike previous flow-based methods that adopt the pre-warping operation sacrificing the original image information as shown in Fig.~\ref{fig:pre_warping}, our FGST combines motion cues with self-attention calculation. Thus, the original image information can be preserved and the guidance effect of the optical flow can be further explored. In addition, our flow-guided scheme enjoys higher flexibility and robustness because adjacent FGABs sample contents independently. Please refer to the SM for detailed discussions.

	\subsection{Recurrent Embedding}
	Our FGSW-MSA is calculated within a short temporal sequence for the computational complexity consideration (approximately linear to the temporal radius $r$ in Eq.~\eqref{eq:complexity_2} ).  Therefore, the receptive field of FGSW-MSA is temporally local and overlooking the distant frames limits the video deblurring  performance. To further capture more robust long-range temporal dependencies, we propose Recurrent Embedding (RE) mechanism. RE is motivated by Recurrent Neural Network (RNN). More specifically, as shown in Fig.~\ref{fig:pipeline} (c), we exploit RE in each Transformer layer to transfer information from past frames and establish long-range temporal correlations. With RE, the FGAB is calculated in a recurrent manner for $T$ time steps. $\boldsymbol{y}^l_t$, $\boldsymbol{e}^l_t, \boldsymbol{q}^l_t$, $\boldsymbol{k}^l_t$ respectively denote the output, RE, $query$ elements, and $key$ elements of the $l_{th}$ FGAB in the $t_{th}$ time step. We have
	\vspace{-1.5mm}
	\begin{equation}
	\small
	\begin{aligned}
	\boldsymbol{e}_t^l &= f_w(\boldsymbol{y}_{t-1}^{l}), 
	~~\boldsymbol{q}_t^l = f_c([\boldsymbol{e}_t^l, \boldsymbol{y}_t^{l-1}]), \\
	\boldsymbol{k}_t^l &= \underset{\footnotesize |j-t| \leq r}{\bigcup~~\boldsymbol{y}_j^{l-1}}, 
	~~\boldsymbol{y}_t^l = \text{FGAB}(\boldsymbol{q}_t^l, \boldsymbol{k}_t^l), \\
	\label{eq:RE_1}
	\end{aligned}
	\vspace{-7.5mm}
	\end{equation}
	where $f_w(\cdot)$ represents the spatial warping that align the feature map at $t$ and ${t-1}$ time step, [$\cdot$,$\cdot$] is the concatenating operation, $f_c(\cdot)$ denotes 3$\times$3 convolution to aggregate the recurrent embedding $\boldsymbol{e}_t^l$ and the output from last FGAB layer $\boldsymbol{y}_t^{l-1}$, and $\boldsymbol{y}_t^l  = \text{FGAB}(\boldsymbol{q}_t^l, \boldsymbol{k}_t^l)$ is formulated in details as
	\vspace{-1.5mm}
	\begin{equation}
	\small
	\begin{aligned}
	\boldsymbol{o}_t^l &= \text{FGSW-MSA}(\text{LN}(\boldsymbol{q}_t^l), \text{LN}(\boldsymbol{k}_t^l) )+ \boldsymbol{q}_t^l, \\
	\boldsymbol{y}_t^l &= \text{FFN}(\boldsymbol{o}_t^l)+\boldsymbol{o}_t^l, \\
	\label{eq:RE_2}
	\end{aligned}
	\vspace{-5.5mm}
	\end{equation}
	where LN denotes the layer normalization and FFN refers to the Feed Forward Network. Our RE sequentially propagates the information from the first frame to the last frame, thus capturing reliable long-range temporal dependencies.
	
	\section{Experiment}
	\vspace{0mm}
	\subsection{Datasets}
	\textbf{DVD.} The DVD~\cite{Su} dataset consists of 71 videos with 6,708 blurry-sharp image pairs. It is divided into \texttt{train/test} subsets with 61 videos (5,708 image pairs) and 10 videos (1,000 image pairs). DVD is captured with mobile phones and DSLR at a frame rate of 240 fps.
	
	\noindent\textbf{GOPRO.} The GOPRO~\cite{GoPro} benchmark is composed of over 3,300 blurry-sharp image pairs of dynamic scenes. It is obtained by a high-speed camera. The training and testing subsets are split in proportional to 2:1. 
	
	\noindent\textbf{Real Blurry Videos.} To validate the generality of FGST, we evaluate models on real blurry datasets collected by \cite{real_blur}. Because the ground truth (GT) is inaccessible, we only compare visual results of FGST and others.
	
	\vspace{-2mm}
	\subsection{Implementation Details}
	We implement FGST in PyTorch. We adopt a pre-trained SPyNet~\cite{spynet} as the optical flow estimator. All the modules are trained with the Adam~\cite{adam} optimizer ($\beta_1$ = 0.9 and $\beta_2$ = 0.999) for 600 epochs. The initial learning rate is set to 2$\times$10$^{-4}$ and 2.5$\times$10$^{-5}$ respectively for the deblurring model and optical flow estimator. The learning rate is halved every 200 epochs during the training procedure. Patches at the size of  256$\times$256 cropped from training frames are fed into the models. The batch size is 8. The temporal radius $r$ of the neighboring frames is set to 1. The sequence length $T$ is set to 9 in training and the whole video length in testing. The horizontal and vertical flips are performed for data augmentation. Peak signal-to-noise ratio (PSNR) and structural similarity (SSIM)~\cite{ssim} are adopted as the evaluation metrics. The models are trained with 8 V100 GPUs. $\mathcal{L}_1$ loss between the restored and GT videos is used for supervision.
	
	\begin{figure*}[t]
		\begin{center}
			\begin{tabular}[t]{c} \hspace{-2mm}
				\includegraphics[width=0.98\textwidth]{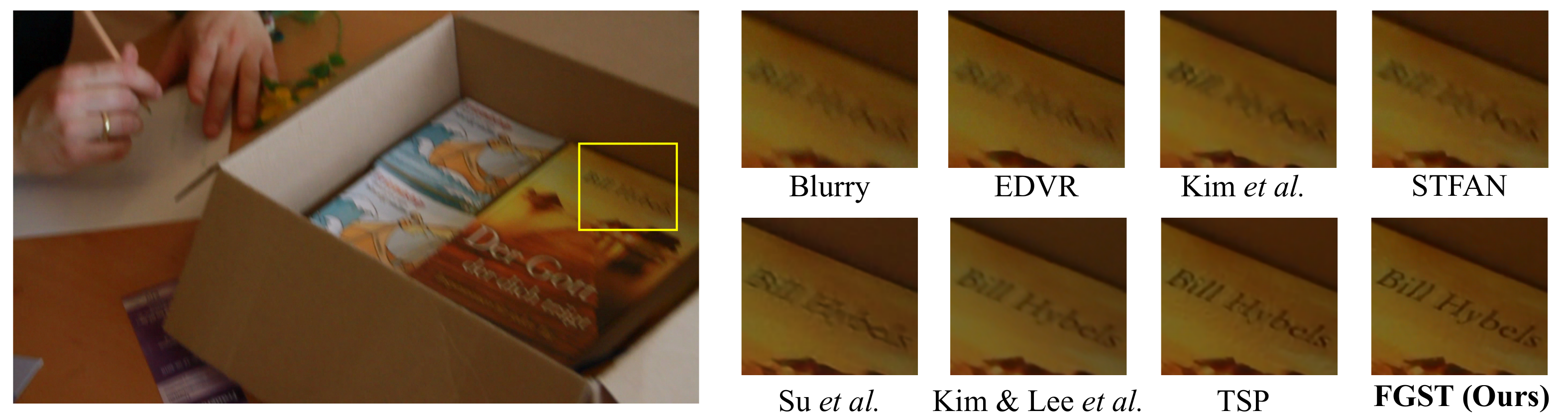}
			\end{tabular}
		\end{center}
		\vspace*{-5mm}
		\caption{\small Visual results of FGST and SOTA methods on the real blurry videos of  \cite{real_blur}. Please zoom in for a better view.}
		\label{fig:real}
		\vspace{-2mm}
	\end{figure*}
	
	\subsection{Quantitative Results}
	\vspace{-1mm}
	The comparisons between FGST and other SOTA methods are listed in Tabs.~\ref{tab:dvd}, \ref{tab:gopro}, and \ref{tab:efficiency}. As can be observed: \textbf{(i)} Our FGST outperforms SOTA methods by a large margin on the two benchmarks. Specifically, as shown in Tab.~\ref{tab:dvd}, our FGST surpasses the recent best algorithm ARVo~\cite{arvo} by 0.56 dB on DVD. As reported in Tab.~\ref{tab:gopro}, our method exceeds Suin \emph{et al.}~\cite{Suin} and TSP~\cite{tsp} by 0.80 dB and 1.23 dB respectively on GOPRO. These results demonstrate the effectiveness of our method. \textbf{(ii)} Tab.~\ref{tab:efficiency} exhibits efficiency comparisons of different algorithms on GOPRO. The FLOPS is tested at the input size of 1$\times$3$\times$240$\times$240. The running time per frame is tested at the spatial size of 1,280$\times$720 on the same RTX 2080 GPU. Our FGST is more cost-effective and achieves a better trade-off between PSNR, Params, FLOPS, and inference speed. For instance, when compared to TSP~\cite{tsp}, FGST only requires 59.9\% (9.70 / 16.19) Params and 36.8\% (131.6 / 357.9) FLOPS while achieving even 1.23 dB improvement and 2.34$\times$ (579.7 / 247.8) speed. This evidence suggests the promising efficiency advantage of our proposed FGST.
	
	\vspace{-1mm}
	\subsection{Qualitative Results}
	\vspace{-1mm}
	We provide visual comparisons on DVD, GOPRO, and real blurry videos as shown in Figs.~\ref{fig:dvd}, \ref{fig:gopro}, and \ref{fig:real}. Previous methods are less favorable to restore abrupt motion blur. They either yield over-smoothing images sacrificing fine textural details and structural contents or introduce redundant blotchy texture and chromatic artifacts when fast motions exists. In contrast, our FGST excels at modeling long-range dependencies and exploits motion information to guide the self-attention module to capture non-local self-similarity in spatio-temporal neighborhoods. As a result, FGST is capable of restoring  structural contents and textural details while preserving spatial smoothness of the homogeneous regions. Supplementary file provides more visual results.
	
	\begin{table*}[t]\vspace{-3mm}
		\subfloat[Break-down ablation study toward better performance. \label{tab:breakdown}]{
			\tablestyle{4pt}{1.05}\scalebox{0.67}{
				\begin{tabular}{c c c  c c}
					\toprule
					\rowcolor{color3}~~Baseline~~ &~~RE~~  &~~FGSW-MSA~~~~~~ &~~~~PSNR~~$\textcolor{black}{\uparrow}$~~~ &~~~~SSIM~~$\textcolor{black}{\uparrow}$~~~\\
					\midrule
					$\checkmark$ & & &31.18~\colorbox{color3}{(+0.00\%)} &0.924~\colorbox{color3}{(+0.00\%)}\\
					$\checkmark$  &$\checkmark$ & &32.34~\colorbox{color3}{(+3.72\%)} &0.943~\colorbox{color3}{(+2.06\%)}\\
					$\checkmark$ & &$\checkmark$ &32.84~\colorbox{color3}{(+5.32\%)} &0.957~\colorbox{color3}{(+3.57\%)}\\
					$\checkmark$  &$\checkmark$ &$\checkmark$ &\bf 32.90~\colorbox{color3}{(+5.52\%)} &\bf 0.961~\colorbox{color3}{(+4.00\%)}\\
					\bottomrule
		\end{tabular}}}\hspace{3.7mm}\vspace{-1mm}
		\subfloat[\small Ablation study of using different self-attention mechanisms.\label{tab:attention}]{
			\tablestyle{4pt}{1.05}\scalebox{0.78}{
				\begin{tabular}{l c c c c c}
					\toprule
					\rowcolor{color3} Method & ~~Baseline~~ & ~~Global MSA~~ &~~Local W-MSA~~   & ~~FGS-MSA~~ & ~~FGSW-MSA~~   \\
					\midrule
					PSNR &31.18 &29.20 &31.71  &\bf 32.48 &\bf 32.84\\
					SSIM &0.924 &0.880 &0.938  &\bf 0.944 &\bf 0.957\\
					Params  &5.15 &64.40 &8.26  &9.69 &9.70\\
					FLOPS  &43.93 &138.68 &108.09 &125.08 &125.67\\
					\bottomrule
		\end{tabular}}}\vspace{-6mm}
		\subfloat[\small Efficiency comparisons with SOTA CNN-based methods. \label{tab:efficiency}]{
			\tablestyle{2.5pt}{1.05}\scalebox{0.78}{\begin{tabular}{l c c c c c}
					\toprule
					\rowcolor{color3}Method &~~~~EDVR~~~~ &~~~Su \emph{et al.}~~~   &~~~STFAN~~~ &~~~~TSP~~~~ &~FGST (Ours)~   \\
					\midrule
					PSNR &26.83 & 27.31 & 28.59 &31.67 &32.90 \\
					Params (M)  &23.60  &15.30  &5.37  &16.19  &9.70\\
					FLOPS (G) &159.2  &38.7  &35.4   &357.9  &131.6\\
					Time (ms/f) &268.5 &133.2 &145.9 &579.7 &247.8\\
					\bottomrule
		\end{tabular}}}\hspace{3.2mm}
		\subfloat[\small FGSW-MSA \emph{v.s.} FGDeConv and DeConv on GOPRO dataset. \label{tab:deformable}]{
			\tablestyle{2.5pt}{1.05}\scalebox{0.78}{\begin{tabular}{l c c c c}
					\toprule
					\rowcolor{color3}Method~~~~ & ~~~~Baseline~~~~ & ~~~~+ DeConv~~~~   &~~~~+ FGDeConv~~~~ &~+ FGSW-MSA~   \\
					\midrule
					PSNR &31.18 &32.35 &32.59 &\bf 32.84\\
					SSIM &0.924 &0.941 &0.954 &\bf 0.957 \\
					Params (M) &5.15 &8.34 &9.78 &9.70\\
					FLOPS (G) &43.93 &108.38 &125.96 &125.67\\
					\bottomrule
		\end{tabular}}}\vspace{-6.5mm}
		\subfloat[\small Pre-warping \emph{v.s.} our FGSW-MSA. \label{tab:model_size}]{
			\tablestyle{2.5pt}{1.05}\scalebox{0.73}{\begin{tabular}{l c c c}
					\toprule
					\rowcolor{color3}Method~ &~Local W-MSA~ &~pre-warping~  &~FGSW-MSA~   \\
					\midrule
					PSNR &31.71  &32.54   &\textbf{32.84}\\
					SSIM &0.938  &0.953   &\textbf{0.957}\\
					\bottomrule
		\end{tabular}}}\hspace{3.7mm}
		\subfloat[\small Ablation study of window sizes.\label{tab:win_size}]{
			\tablestyle{4.8pt}{1.05}\scalebox{0.73}{
				\begin{tabular}{c c c c c c }
					\toprule
					\rowcolor{color3} ~Win Size~ &~1$\times$1~ &~2$\times$2~  &~3$\times$3~ &~4$\times$4~ &~5$\times$5~  \\
					\midrule
					PSNR  &32.48 &32.62 &\bf 32.90 &32.71 &32.66 \\
					SSIM  &0.944 &0.955 &\bf 0.961  &0.955 &0.957  \\
					\bottomrule
		\end{tabular}}}\hspace{3.7mm}
		\subfloat[\small Ablation study of optical flow estimators.\label{tab:flow}]{
			\tablestyle{2.2pt}{1.05}\scalebox{0.73}{
				\begin{tabular}{l c c c c c}
					\toprule
					\rowcolor{color3}Method~&~~Baseline~~ &~~FlowNet~~  &~~SPyNet~~  &~~PWC-Net~~  \\
					\midrule
					PSNR~$\textcolor{black}{\uparrow}$  &31.18 &32.85 &32.90 &\bf 33.03\\
					SSIM~~$\textcolor{black}{\uparrow}$  &0.924 &0.960 &0.961 &\bf 0.964\\
					\bottomrule
		\end{tabular}}}\vspace{-3mm}
		\caption{\small Ablation studies. The models are trained and tested on GOPRO. PSNR, SSIM, Params, FLOPS, and inference time  are reported.}
		\label{tab:ablations}\vspace{-5mm}
	\end{table*}
	
	\vspace{-10mm}
	\subsection{Ablation Study}
	\vspace{-1mm}
	In this part, we conduct ablation studies on GOPRO dataset. The baseline model is derived by directly removing all the proposed RE and FGSW-MSA modules from our FGST. 
	
	\noindent\textbf{Break-down Ablation.} We firstly conduct a break-down ablation to investigate the effect of each component toward better performance. The results are reported in Tab.~\ref{tab:breakdown}. The baseline model yields 31.18 dB. After applying RE and FGSW-MSA respectively, the deblurring model achieves 1.16 dB and 1.66 dB improvements. While using both RE and FGSW-MSA modules, the model gains by 1.72 dB. The results suggest the effectiveness of RE and FGSW-MSA.
	
	\noindent\textbf{Self-Attention Mechanism.} We compare our self-attention mechanisms with other competitors in Tab.~\ref{tab:attention}. The baseline model yields 31.18 dB while costing 5.15M Params and 43.93G FLOPS. \textbf{(i)} When using global MSA~\cite{global_msa}, the feature maps are downsampled into $\frac{1}{4}$ size and the channel is increased by 4 times to avoid out of memory and information loss. The deblurring model degrades by 1.98 dB while costing 12.5$\times$ Params and 3.2$\times$ FLOPS. This is mainly because global MSA attends to too redundant $key$ elements, requiring a large amount of computation and memory resources while leading to ambiguous gradients for input features~\cite{de_detr} and thus non-convergence problem. Meanwhile, features from global aggregation tend to over-smooth the predictions of small patterns~\cite{xiangtl_gald}. \textbf{(ii)} When using local W-MSA~\cite{liu2021swin}, the model gains by only 0.53 dB while adding 3.11M Params and 64.16G FLOPS. The improvement is limited while the additional burden is nontrivial. That is because W-MSA calculates self-attention within position-specific windows. The receptive field is limited. \textbf{(iii)} Our FGS-MSA exploits the optical flow as the guidance to sample spatially sparse $keys$ of similar and sharper regions in the spatio-temporal neighborhood for each $query$ on the reference frame. Compared to global MSA, the $key$ elements of FGST are less but highly related to the selected $query$. Thus, when using FGS-MSA, the model gains by 1.30 dB while adding 4.54M Params and 81.15G FLOPS. These results show that FGS-MSA costs cheaper resources but achieves better performance than global MSA. When exploiting FGSW-MSA, the model yields an improvement of 1.72 dB while adding 4.55M Params and 87.69G FLOPS. This evidence suggests: \textbf{(a)} FGSW-MSA is more effective than W-MSA in fast motion blur restoration. \textbf{(b)} FGSW-MSA is more reliable than FGS-MSA and achieves better deblurring performance.
	
	In addition, we conduct visual analysis on three adjacent frames by visualizing the last feature map of models with and without (w/o) FGSW-MSA in Fig.~\ref{fig:fea}. Deeper color indicates larger weights. It can be observed that the model without FGSW-MSA responds weakly to similar regions in the neighboring frames. In contrast, the model equipped with FGSW-MSA generates much stronger responses to highly related but misaligned scene patches. Moreover, FGST pays more attention to the regions with fast motion blur. These results demonstrates the effectiveness of FGSW-MSA in capturing non-local self-similarity in dynamic scenes.
	
	\noindent\textbf{Flow-Guided Deformable Convolution.} We compare our FGSW-MSA with deformable convolution (DeConv)~\cite{edvr} and recent flow-guided deformable convolution (FGDeConv)~\cite{chan2021basicvsr++} in Tab.~\ref{tab:deformable}. Our proposed FGSW-MSA achieves the most significant improvement. This mainly stems from that FGSW-MSA excels at capturing non-local similarity and long-range dependencies, which are the limitations of CNN-based methods.
	
	\noindent\textbf{Pre-warping Strategy.} We compare our FGSW-MSA with the pre-warping strategy mainly adopted by previous methods in Tab.~\ref{tab:model_size}. We start from the baseline model equipped with W-MSA. It can be observed that using FGSW-MSA is 0.30 dB and 0.004 in terms of PSNR and SSIM higher than using pre-warping operation. This performance gap is mainly because the model using our FGSW-MSA can learn from non-corrupted representations of input video and further  explore the guidance effect of the optical flow.
	
	\noindent\textbf{Window Size.} We change the window size of FGSW-MSA to study its effect. The results are listed in Tab.~\ref{tab:win_size}. We start by setting the window size at 1$\times$1 and then gradually increase it. The performance  achieves its maximum when the window size is 3$\times$3. Thus, the optimal setting is 3$\times$3.

	\noindent\textbf{Optical Flow Estimator.} We adopt three representative optical flow estimators (FlowNet~\cite{flownet}, SPyNet~\cite{spynet}, and PWC-Net~\cite{pwcnet}) to investigate their effects in Tab.~\ref{tab:flow}. \textbf{(i)} No matter what flow estimator is used, FGST reliably outperforms the baseline model, suggesting the robustness and generality of our method. \textbf{(ii)} The performance of FGST can be further improved by using a better flow estimator. To be specific, when equipped with PWC-Net, FGST is 0.18 dB and 0.13 dB higher than those using FlowNet and SPyNet. These results demonstrate that FGST can directly and conveniently enjoy the benefits of SOTA optical flow estimators.
	
	\vspace{-2mm}
	\section{Conclusion}
	\vspace{-1mm}
	In this paper, we propose a novel  Transformer-based method, FGST, for video deblurring. In FGST, we customize a self-attention mechanism, FGS-MSA, and then promote it to FGSW-MSA. Guided by an optical flow estimator, FGSW-MSA samples spatially sparse but highly related $key$ elements corresponding to similar and sharper scene patches in the spatio-temporal neighborhoods. Besides, we present an embedding scheme, RE, to transfer information of past frames and capture long-range temporal dependencies. Comprehensive experiments demonstrate that our FGST significantly surpasses SOTA methods and generates more visually pleasant results in real video deblurring.

	\textbf{Acknowledgements:} This work is partially supported by the NSFC fund (61831014), the Shenzhen Science and Technology Project under Grant (CJGJZD20200617102601004, JSGG20210802153150005).
	
	\bibliography{reference_mst}

\begin{thebibliography}{56}
\providecommand{\natexlab}[1]{#1}
\providecommand{\url}[1]{\texttt{#1}}
\expandafter\ifx\csname urlstyle\endcsname\relax
  \providecommand{\doi}[1]{doi: #1}\else
  \providecommand{\doi}{doi: \begingroup \urlstyle{rm}\Url}\fi

\bibitem[Arnab et~al.(2021)Arnab, Dehghani, Heigold, Sun, Lu{\v{c}}i{\'c}, and
  Schmid]{arnab2021vivit}
Arnab, A., Dehghani, M., Heigold, G., Sun, C., Lu{\v{c}}i{\'c}, M., and Schmid,
  C.
\newblock Vivit: A video vision transformer.
\newblock \emph{arXiv preprint arXiv:2103.15691}, 2021.

\bibitem[Cai et~al.(2020)Cai, Wang, Luo, Yin, Du, Wang, Zhou, Zhou, Zhang, and
  Sun]{rsn}
Cai, Y., Wang, Z., Luo, Z., Yin, B., Du, A., Wang, H., Zhou, X., Zhou, E.,
  Zhang, X., and Sun, J.
\newblock Learning delicate local representations for multi-person pose
  estimation.
\newblock In \emph{ECCV}, 2020.

\bibitem[Cai et~al.(2021{\natexlab{a}})Cai, Hu, Wang, Zhang, Pfister, and
  Wei]{cai2021learning}
Cai, Y., Hu, X., Wang, H., Zhang, Y., Pfister, H., and Wei, D.
\newblock Learning to generate realistic noisy images via pixel-level
  noise-aware adversarial training.
\newblock In \emph{NeurIPS}, 2021{\natexlab{a}}.

\bibitem[Cai et~al.(2021{\natexlab{b}})Cai, Lin, Hu, Wang, Yuan, Zhang,
  Timofte, and Van~Gool]{cai2021mask}
Cai, Y., Lin, J., Hu, X., Wang, H., Yuan, X., Zhang, Y., Timofte, R., and
  Van~Gool, L.
\newblock Mask-guided spectral-wise transformer for efficient hyperspectral
  image reconstruction.
\newblock \emph{arXiv preprint arXiv:2111.07910}, 2021{\natexlab{b}}.

\bibitem[Cao et~al.(2021{\natexlab{a}})Cao, Wang, Chen, Jiang, Zhang, Tian, and
  Wang]{cao2021swin}
Cao, H., Wang, Y., Chen, J., Jiang, D., Zhang, X., Tian, Q., and Wang, M.
\newblock Swin-unet: Unet-like pure transformer for medical image segmentation.
\newblock \emph{arXiv preprint arXiv:2105.05537}, 2021{\natexlab{a}}.

\bibitem[Cao et~al.(2021{\natexlab{b}})Cao, Li, Zhang, and Van~Gool]{vsrt}
Cao, J., Li, Y., Zhang, K., and Van~Gool, L.
\newblock Video super-resolution transformer.
\newblock \emph{arXiv preprint arXiv:2106.06847}, 2021{\natexlab{b}}.

\bibitem[Carion et~al.(2020)Carion, Massa, Synnaeve, Usunier, Kirillov, and
  Zagoruyko]{to_1}
Carion, N., Massa, F., Synnaeve, G., Usunier, N., Kirillov, A., and Zagoruyko,
  S.
\newblock End-to-end object detection with transformers.
\newblock In \emph{ECCV}, 2020.

\bibitem[Chakrabarti(2016)]{chakrabarti2016neural}
Chakrabarti, A.
\newblock A neural approach to blind motion deblurring.
\newblock In \emph{ECCV}, 2016.

\bibitem[Chan et~al.(2021)Chan, Zhou, Xu, and Loy]{chan2021basicvsr++}
Chan, K.~C., Zhou, S., Xu, X., and Loy, C.~C.
\newblock Basicvsr++: Improving video super-resolution with enhanced
  propagation and alignment.
\newblock 2021.

\bibitem[Chen et~al.(2021)Chen, Wang, Guo, Xu, Deng, Liu, Ma, Xu, Xu, and
  Gao]{ipt}
Chen, H., Wang, Y., Guo, T., Xu, C., Deng, Y., Liu, Z., Ma, S., Xu, C., Xu, C.,
  and Gao, W.
\newblock Pre-trained image processing transformer.
\newblock In \emph{CVPR}, 2021.

\bibitem[Cho et~al.(2012)Cho, Wang, Lee, b, and c]{real_blur}
Cho, S., Wang, J., Lee, S., b, and c.
\newblock Video deblurring for hand-held cameras using patch-based synthesis.
\newblock In \emph{ACM TOG}, 2012.

\bibitem[Dosovitskiy et~al.(2015)Dosovitskiy, Fischer, Ilg, H{\"a}usser,
  Haz{\i}rba{\c{s}}, Golkov, v.d. Smagt, Cremers, and Brox]{flownet}
Dosovitskiy, A., Fischer, P., Ilg, E., H{\"a}usser, P., Haz{\i}rba{\c{s}}, C.,
  Golkov, V., v.d. Smagt, P., Cremers, D., and Brox, T.
\newblock Flownet: Learning optical flow with convolutional networks.
\newblock In \emph{ICCV}, 2015.

\bibitem[Dosovitskiy et~al.(2021)Dosovitskiy, Beyer, Kolesnikov, Weissenborn,
  Zhai, Unterthiner, Dehghani, Minderer, Heigold, Gelly, Uszkoreit, and
  Houlsby]{global_msa}
Dosovitskiy, A., Beyer, L., Kolesnikov, A., Weissenborn, D., Zhai, X.,
  Unterthiner, T., Dehghani, M., Minderer, M., Heigold, G., Gelly, S.,
  Uszkoreit, J., and Houlsby, N.
\newblock An image is worth 16x16 words: Transformers for image recognition at
  scale.
\newblock In \emph{ICLR}, 2021.

\bibitem[El-Nouby et~al.(2021)El-Nouby, Touvron, Caron, Bojanowski, Douze,
  Joulin, Laptev, Neverova, Synnaeve, Verbeek, et~al.]{xcit}
El-Nouby, A., Touvron, H., Caron, M., Bojanowski, P., Douze, M., Joulin, A.,
  Laptev, I., Neverova, N., Synnaeve, G., Verbeek, J., et~al.
\newblock Xcit: Cross-covariance image transformers.
\newblock \emph{arXiv preprint arXiv:2106.09681}, 2021.

\bibitem[Gast \& Roth(2019)Gast and Roth]{gast2019deep}
Gast, J. and Roth, S.
\newblock Deep video deblurring: The devil is in the details.
\newblock In \emph{ICCVW}, 2019.

\bibitem[Gong et~al.(2017)Gong, Yang, Liu, Zhang, Reid, Shen, Van Den~Hengel,
  and Shi]{gong2017motion}
Gong, D., Yang, J., Liu, L., Zhang, Y., Reid, I., Shen, C., Van Den~Hengel, A.,
  and Shi, Q.
\newblock From motion blur to motion flow: A deep learning solution for
  removing heterogeneous motion blur.
\newblock In \emph{CVPR}, 2017.

\bibitem[Hu et~al.(2021)Hu, Ma, Liu, Cai, Zhao, Zhang, and Wang]{hu2021pseudo}
Hu, X., Ma, R., Liu, Z., Cai, Y., Zhao, X., Zhang, Y., and Wang, H.
\newblock Pseudo 3d auto-correlation network for real image denoising.
\newblock In \emph{CVPR}, 2021.

\bibitem[Hyun~Kim et~al.(2017)Hyun~Kim, Mu~Lee, Scholkopf, and
  Hirsch]{hyun2017online}
Hyun~Kim, T., Mu~Lee, K., Scholkopf, B., and Hirsch, M.
\newblock Online video deblurring via dynamic temporal blending network.
\newblock In \emph{ICCV}, 2017.

\bibitem[Jin et~al.(2005)Jin, Favaro, Cipolla, b, and c]{track}
Jin, H., Favaro, P., Cipolla, R., b, and c.
\newblock Visual tracking in the presence of motion blur.
\newblock In \emph{CVPR}, 2005.

\bibitem[Kim et~al.(2015)Kim, Mu~Lee, a, and b]{Kim}
Kim, H., Mu~Lee, K., a, and b.
\newblock Generalized video deblurring for dynamic scenes.
\newblock In \emph{CVPR}, 2015.

\bibitem[Kingma \& Ba(2015)Kingma and Ba]{adam}
Kingma, D.~P. and Ba, J.~L.
\newblock Adam: A method for stochastic optimization.
\newblock In \emph{ICLR}, 2015.

\bibitem[Li et~al.(2021{\natexlab{a}})Li, Xu, Zhang, Yu, Zhong, Ren, Suominen,
  and Li]{arvo}
Li, D., Xu, C., Zhang, K., Yu, X., Zhong, Y., Ren, W., Suominen, H., and Li, H.
\newblock Arvo: Learning all-range volumetric correspondence for video
  deblurring.
\newblock In \emph{CVPR}, 2021{\natexlab{a}}.

\bibitem[Li et~al.(2021{\natexlab{b}})Li, Wang, Zhang, Xu, Xu, and Tu]{prtr}
Li, K., Wang, S., Zhang, X., Xu, Y., Xu, W., and Tu, Z.
\newblock Pose recognition with cascade transformers.
\newblock In \emph{CVPR}, 2021{\natexlab{b}}.

\bibitem[Li et~al.(2019)Li, Zhang, You, Yang, Yang, and Tong]{xiangtl_gald}
Li, X., Zhang, L., You, A., Yang, M., Yang, K., and Tong, Y.
\newblock Global aggregation then local distribution in fully convolutional
  networks.
\newblock In \emph{BMVC}, 2019.

\bibitem[Li et~al.(2010)Li, Kang, Joshi, Seitz, and
  Huttenlocher]{li2010generating}
Li, Y., Kang, S.~B., Joshi, N., Seitz, S.~M., and Huttenlocher, D.~P.
\newblock Generating sharp panoramas from motion-blurred videos.
\newblock In \emph{CVPR}, 2010.

\bibitem[Liu et~al.(2021)Liu, Lin, Cao, Hu, Wei, Zhang, Lin, and
  Guo]{liu2021swin}
Liu, Z., Lin, Y., Cao, Y., Hu, H., Wei, Y., Zhang, Z., Lin, S., and Guo, B.
\newblock Swin transformer: Hierarchical vision transformer using shifted
  windows.
\newblock In \emph{ICCV}, 2021.

\bibitem[Makansi et~al.(2017)Makansi, Ilg, Brox, b, and c]{makansi2017end}
Makansi, O., Ilg, E., Brox, T., b, and c.
\newblock End-to-end learning of video super-resolution with motion
  compensation.
\newblock In \emph{GCPR}, 2017.

\bibitem[Matsushita et~al.(2006)Matsushita, Ofek, Ge, Tang, and
  Shum]{matsushita2006full}
Matsushita, Y., Ofek, E., Ge, W., Tang, X., and Shum, H.-Y.
\newblock Full-frame video stabilization with motion inpainting.
\newblock \emph{TPAMI}, 2006.

\bibitem[Nah et~al.(2017)Nah, Hyun~Kim, Mu~Lee, and b]{GoPro}
Nah, S., Hyun~Kim, T., Mu~Lee, K., and b.
\newblock Deep multi-scale convolutional neural network for dynamic scene
  deblurring.
\newblock In \emph{CVPR}, 2017.

\bibitem[Nah et~al.(2019)Nah, Son, Lee, and b]{Nah}
Nah, S., Son, S., Lee, K.~M., and b.
\newblock Recurrent neural networks with intra-frame iterations for video
  deblurring.
\newblock In \emph{CVPR}, 2019.

\bibitem[Pan et~al.(2020)Pan, Bai, Tang, and b]{tsp}
Pan, J., Bai, H., Tang, J., and b.
\newblock Cascaded deep video deblurring using temporal sharpness prior.
\newblock In \emph{CVPR}, 2020.

\bibitem[Purohit et~al.(2020)Purohit, Rajagopalan, b, and c]{purohit2020region}
Purohit, K., Rajagopalan, A., b, and c.
\newblock Region-adaptive dense network for efficient motion deblurring.
\newblock In \emph{AAAI}, 2020.

\bibitem[Ramachandran et~al.(2019)Ramachandran, Parmar, Vaswani, Bello,
  Levskaya, and Shlens]{tc_2}
Ramachandran, P., Parmar, N., Vaswani, A., Bello, I., Levskaya, A., and Shlens,
  J.
\newblock Stand-alone self-attention in vision models.
\newblock In \emph{NeurIPS}, 2019.

\bibitem[Ranjan et~al.(2017)Ranjan, Black, b, and c]{spynet}
Ranjan, A., Black, M.~J., b, and c.
\newblock Optical flow estimation using a spatial pyramid network.
\newblock In \emph{CVPR}, 2017.

\bibitem[Ronneberger et~al.(2015)Ronneberger, Fischer, Brox, a, and b]{unet}
Ronneberger, O., Fischer, P., Brox, T., a, and b.
\newblock U-net: Convolutional networks for biomedical image segmentation.
\newblock In \emph{MICCAI}, 2015.

\bibitem[Su et~al.(2017)Su, Delbracio, Wang, Sapiro, Heidrich, and Wang]{Su}
Su, S., Delbracio, M., Wang, J., Sapiro, G., Heidrich, W., and Wang, O.
\newblock Deep video deblurring for hand-held cameras.
\newblock In \emph{CVPR}, 2017.

\bibitem[Suin et~al.(2021)Suin, Rajagopalan, b, and c]{Suin}
Suin, M., Rajagopalan, A.~N., b, and c.
\newblock Gated spatio-temporal attention-guided video deblurring.
\newblock In \emph{CVPR}, 2021.

\bibitem[Sun et~al.(2018)Sun, Yang, Liu, and Kautz]{pwcnet}
Sun, D., Yang, X., Liu, M.-Y., and Kautz, J.
\newblock {PWC-Net}: {CNNs} for optical flow using pyramid, warping, and cost
  volume.
\newblock In \emph{CVPR}, 2018.

\bibitem[Sun et~al.(2015)Sun, Cao, Xu, and Ponce]{sun2015learning}
Sun, J., Cao, W., Xu, Z., and Ponce, J.
\newblock Learning a convolutional neural network for non-uniform motion blur
  removal.
\newblock In \emph{CVPR}, 2015.

\bibitem[Tao et~al.(2018)Tao, Gao, Shen, Wang, and Jia]{Tao}
Tao, X., Gao, H., Shen, X., Wang, J., and Jia, J.
\newblock Scale-recurrent network for deep image deblurring.
\newblock In \emph{CVPR}, 2018.

\bibitem[Vaswani et~al.(2017)Vaswani, Shazeer, Parmar, Uszkoreit, Jones, Gomez,
  Kaiser, and Polosukhin]{vaswani2017attention}
Vaswani, A., Shazeer, N., Parmar, N., Uszkoreit, J., Jones, L., Gomez, A.~N.,
  Kaiser, {\L}., and Polosukhin, I.
\newblock Attention is all you need.
\newblock In \emph{NeurIPS}, 2017.

\bibitem[Vaswani et~al.(2021)Vaswani, Ramachandran, Srinivas, Parmar, Hechtman,
  and Shlens]{vaswani2021scaling}
Vaswani, A., Ramachandran, P., Srinivas, A., Parmar, N., Hechtman, B., and
  Shlens, J.
\newblock Scaling local self-attention for parameter efficient visual
  backbones.
\newblock In \emph{CVPR}, 2021.

\bibitem[Wang et~al.(2019)Wang, Chan, Yu, Dong, and Change~Loy]{edvr}
Wang, X., Chan, K.~C., Yu, K., Dong, C., and Change~Loy, C.
\newblock Edvr: Video restoration with enhanced deformable convolutional
  networks.
\newblock In \emph{CVPRW}, 2019.

\bibitem[Wang et~al.(2004)Wang, Bovik, Sheikh, and Simoncell]{ssim}
Wang, Z., Bovik, A.~C., Sheikh, H.~R., and Simoncell, E.~P.
\newblock Image quality assessment: from error visibility to structural
  similarity.
\newblock \emph{TIP}, 2004.

\bibitem[Wang et~al.(2021)Wang, Cun, Bao, and Liu]{uformer}
Wang, Z., Cun, X., Bao, J., and Liu, J.
\newblock Uformer: A general u-shaped transformer for image restoration.
\newblock \emph{arXiv preprint 2106.03106}, 2021.

\bibitem[Wu et~al.(2020)Wu, Xu, Dai, Wan, Zhang, Yan, Tomizuka, Gonzalez,
  Keutzer, and Vajda]{tc_3}
Wu, B., Xu, C., Dai, X., Wan, A., Zhang, P., Yan, Z., Tomizuka, M., Gonzalez,
  J., Keutzer, K., and Vajda, P.
\newblock Visual transformers: Token-based image representation and processing
  for computer vision.
\newblock \emph{arXiv preprint arXiv:2006.03677}, 2020.

\bibitem[Xiang et~al.(2020)Xiang, Wei, Wai, and Pan]{Xiang}
Xiang, X., Wei, H., Wai, H., and Pan, J.
\newblock Deep video deblurring using sharpness features from exemplars.
\newblock In \emph{TIP}, 2020.

\bibitem[Xue et~al.(2019)Xue, Chen, Wu, Wei, and Freeman]{xue2019video}
Xue, T., Chen, B., Wu, J., Wei, D., and Freeman, W.~T.
\newblock Video enhancement with task-oriented flow.
\newblock \emph{IJCV}, 2019.

\bibitem[Yin et~al.(2021)Yin, Zhou, Krähenbühl, b, and c]{3d_det}
Yin, T., Zhou, X., Krähenbühl, P., b, and c.
\newblock Center-based 3d object detection and tracking.
\newblock In \emph{CVPR}, 2021.

\bibitem[Zhang et~al.(2013)Zhang, Wipf, and Zhang]{zhang2013multi}
Zhang, H., Wipf, D., and Zhang, Y.
\newblock Multi-image blind deblurring using a coupled adaptive sparse prior.
\newblock In \emph{CVPR}, 2013.

\bibitem[Zhang et~al.(2018)Zhang, Luo, Zhong, Lin~Ma, and Li]{dblrnet}
Zhang, K., Luo, W., Zhong, Y., Lin~Ma, W.~L., and Li, H.
\newblock Adversarial spatio-temporal learning for video deblurring.
\newblock In \emph{TIP}, 2018.

\bibitem[Zheng et~al.(2021)Zheng, Lu, Zhao, Zhu, Luo, Wang, Fu, Feng, Xiang,
  Torr, and Zhang]{SETR}
Zheng, S., Lu, J., Zhao, H., Zhu, X., Luo, Z., Wang, Y., Fu, Y., Feng, J.,
  Xiang, T., Torr, P.~H., and Zhang, L.
\newblock Rethinking semantic segmentation from a sequence-to-sequence
  perspective with transformers.
\newblock In \emph{CVPR}, 2021.

\bibitem[Zhong et~al.(2020)Zhong, Gao, Zheng, and Zheng]{RNN_3}
Zhong, Z., Gao, Y., Zheng, Y., and Zheng, B.
\newblock Efficient spatio-temporal recurrent neural network for video
  deblurring.
\newblock In \emph{ECCV}, 2020.

\bibitem[Zhou et~al.(2019)Zhou, Zhang, Pan, Xie, Zuo, and Ren]{stfan}
Zhou, S., Zhang, J., Pan, J., Xie, H., Zuo, W., and Ren, J.
\newblock Spatio-temporal filter adaptive network for video deblurring.
\newblock In \emph{ICCV}, 2019.

\bibitem[Zhu et~al.(2020)Zhu, Su, Lu, Li, Wang, and Dai]{de_detr}
Zhu, X., Su, W., Lu, L., Li, B., Wang, X., and Dai, J.
\newblock Deformable detr: Deformable transformers for end-to-end object
  detection.
\newblock \emph{arXiv preprint arXiv:2010.04159}, 2020.

\bibitem[Zoran et~al.(2011)Zoran, Weiss, b, and c]{zoran2011learning}
Zoran, D., Weiss, Y., b, and c.
\newblock From learning models of natural image patches to whole image
  restoration.
\newblock In \emph{ICCV}, 2011.

\end{thebibliography}
	\bibliographystyle{icml2022}
\end{document}